\documentclass[conference]{IEEEtran}
\IEEEoverridecommandlockouts

\usepackage{cite}
\usepackage{amsmath,amssymb,amsfonts}
\usepackage{algorithmic}
\usepackage{graphicx}
\usepackage{textcomp}
\usepackage{xcolor}
\usepackage{svg}
\usepackage{pdfpages}
\usepackage{multirow}
\usepackage{booktabs}
\usepackage{hyperref}
\usepackage{cleveref}

\def\BibTeX{{\rm B\kern-.05em{\sc i\kern-.025em b}\kern-.08em
    T\kern-.1667em\lower.7ex\hbox{E}\kern-.125emX}}

\begin{document}

\title{Smooth-Foley: Creating Continuous Sound for Video-to-Audio Generation Under Semantic Guidance\\

}

\author{
\IEEEauthorblockN{Yaoyun Zhang, Xuenan Xu, Mengyue Wu}
\IEEEauthorblockA{
    \textit{MoE Key Lab of Artificial Intelligence, AI Institute}\\
    \textit{
     X-LANCE Lab, Department of Computer Science and Engineering} \\
     \textit{Shanghai Jiao Tong University
     }\\
Shanghai, China \\
\{dancloud, wsntxxn, mengyuewu\}@sjtu.edu.cn}

}
\maketitle

\begin{abstract}
The video-to-audio (V2A) generation task has drawn attention in the field of multimedia due to the practicality in producing Foley sound.
Semantic and temporal conditions are fed to the generation model to indicate sound events and temporal occurrence.
Recent studies on synthesizing immersive and synchronized audio are faced with challenges on videos with moving visual presence.
The temporal condition is not accurate enough while low-resolution semantic condition exacerbates the problem.
To tackle these challenges, we propose Smooth-Foley, a V2A generative model taking semantic guidance from the textual label across the generation to enhance both semantic and temporal alignment in audio.
Two adapters are trained to leverage pre-trained text-to-audio generation models.
A frame adapter integrates high-resolution frame-wise video features while a temporal adapter integrates temporal conditions obtained from similarities of visual frames and textual labels.
The incorporation of semantic guidance from textual labels achieves precise audio-video alignment.
We conduct extensive quantitative and qualitative experiments.
Results show that Smooth-Foley performs better than existing models on both continuous sound scenarios and general scenarios.
With semantic guidance, the audio generated by Smooth-Foley exhibits higher quality and better adherence to physical laws.
\end{abstract}


\section{Introduction}
\label{sec:intro}
Recent advances in video-to-audio (V2A) generation models have promoted the development of AI-generative contents, especially Foley in film and video post-processing.
Since creating high-quality audio with precise, continuous synchronization requires specialty and is labor-intensive, automating the Foley process with tools is highly anticipated.
\begin{figure*}[t]
\centering
\includegraphics[width=\textwidth]{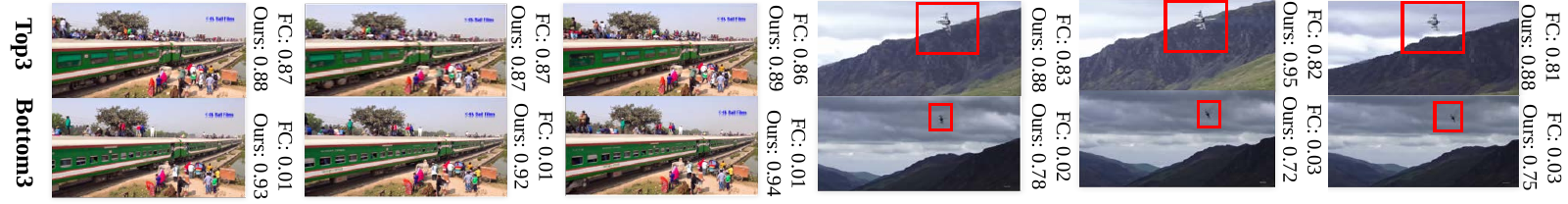}
\caption{\textbf{Examples to clarify the deficits.} In the context of continuous sound and ambiguous object, FoleyCrafter (FC) fails to predict the presence of sound. In the first case, a train stays in the video for a long time but FC predicts event probabilities of nearly zero for some frames. In the second case, when flying airplane becomes tiny, FC fails to detect it, leading to unsatisfactory generation results.}
\label{fig:foleycrafter_deficits}
\end{figure*}

Two objectives are crucial in V2A generation: 1) semantic alignment: the generated sound events should be consistent with the video content; 2) temporal alignment: the generated sound should be synchronized with video frames.
Existing V2A works endeavored to improve the generation performance from two directions.
One direction is to employ increasingly advanced generation models.
Pioneering V2A models were based on generative adversarial networks (GAN) \cite{chen2020generating} or auto-regressive models~\cite{iashin2021taming,du2023conditional}.
Subsequent works employed diffusion models~\cite{ho2020denoising} or flow matching~\cite{wang2024frieren} to further advance the generation quality.
Another direction is to improve the generation quality and controllability by incorporating various conditions.
The condition can be semantic and temporal-relevant video embeddings.
For example, Diff-Foley~\cite{luo2024diff} took video features from contrastive pre-training on aligned video-audio data as a better condition.
Some works also provided explicit signals with physical meanings as conditions, such as audio timbre prompt~\cite{du2023conditional,cui2023varietysound}, temporal conditions~\cite{zhang2024foleycrafter} and energy conditions~\cite{lee2024video}.

Although existing methods have exhibited better temporal alignment in V2A generation, they retain certain limitations.
Specifically, previous models cannot generate continuous, long-duration sound for videos characterized with moving visual presence, \textit{e.g., flying aircraft and off-screen audible siren}.
Typical examples are shown in \Cref{fig:foleycrafter_deficits}.
This indicates one aspect of the insufficiency guidance: \textbf{the temporal condition is not accurate enough}. 
Another aspect stems from \textbf{the low temporal resolution of semantic video condition}.
For example, the resolution of video features in Diff-Foley was 4fps, far smaller than 30fps in common videos.
The low temporal resolution of video features leads to a rough synchronization between audio and video, influencing the temporal alignment performance. 


In this work, we propose Smooth-Foley to achieve smooth and continuous V2A generation under semantic guidance.
The semantic guidance improves conditions by involving textual labels and finer frame-level video embeddings.
We follow the architecture of FoleyCrafter~\cite{zhang2024foleycrafter}, which adapts pre-trained text-to-audio (T2A) generation model for V2A using efficient adapters. 
First, we improve the accuracy of temporal condition by utilizing the label as an additional guidance for semantic consistency.
CLIP~\cite{radford2021learning} similarities between video frames and the label are taken as the temporal condition.
Second, high-resolution frame-wise video embeddings are fed to the generation model to enhance the effectiveness of semantic conditions.
By infusing these two methods, our model not only achieves better performance on continuous sound categories but also lead to more temporally-aligned V2A generation.

Contributions are summarized as: 
1) We integrate frame-wise video features to enhance the temporal resolution of semantic conditions, thereby enhancing the realistic and immersive sound effects.
2) We enhance the temporal condition with the guidance of textual label. By integrating CLIP and textual labels, the temporal alignment between generated audio and video is improved.
3) By efficient fine-tuning of a pre-trained T2A model, Smooth-Foley exhibits increased performance on VGGSound, demonstrating control over continuous sound and understanding of physical laws.



\begin{figure}[htp]
\centering
\includegraphics[width=0.5\textwidth]{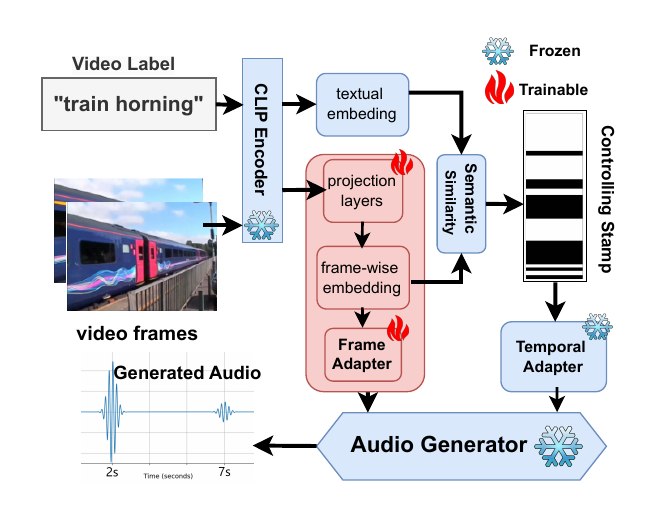}
\caption{\textbf{Overall pipeline of Smooth-Foley.} Note that frame adapter and temporal controller module are trained separately. }
\label{fig:architecture}
\end{figure}

\section{Smooth-Foley}
As previously described, Smooth-Foley integrates pre-trained T2A models by lightweight adapters.
Auffusion~\cite{xue2024auffusion} is chosen as the T2A model, enabling adaptation to data-scarcity scenarios while keeping high-fidelity and diverse audio synthesis abilities. 
As shown in \Cref{fig:architecture}, conditions fed to the generation model are from two modules: a frame adapter and a temporal adapter. 
The two adapters are trained separately.
When training one adapter, all other modules are kept frozen.
The video label is incorporated to enhance the accuracy of temporal conditions.
We first introduce the semantic guidance to improve the performance and then elaborate on the two modules.
Finally, we describe the data filtered to support efficient fine-tuning.

\subsection{Semantic Guidance}
In Smooth-Foley, semantic guidance stems from two aspects: 1) we adopt frame-wise video guidance instead of clip-wise guidance to enhance the granularity of visual conditions. Since frames inherently carry temporal information, this process enhances both temporal and semantic alignment;
2) textual label is utilized for more accurate and coherent temporal conditions.
Since a video may contain multiple objects, the label serves as important guidance to detect the occurrence of the sounding object, providing more accurate predictions than those solely from visual frames.

\subsection{Frame Adapter with Frame-Wise Visual Guidance}
\subsubsection{Visual Encoder}
Though CLIP encoder has shown effectiveness in extracting visual semantic features, we need to adapt it for V2A generation.
Therefore, we adopt an adapter to project frame features from CLIP, formatted as:
\begin{equation}
    V_{frame} = MLP(\mathcal{E}_{clip}(v))
\end{equation}
where $v$ is the input video frames, $\mathcal{E}_{clip}$ represents the frozen CLIP image encoder, and $MLP$ denotes a learnable projection module.
We follow the settings from IP-Adapter~\cite{ye2023ip}, to use a linear projection and feed frame embeddings into the frozen T2A model.

\subsubsection{Frame Adapter}
Following FoleyCrafter, we integrate visual features and textual features with the frozen T2A backbone by parallel cross-attention adapters.
Instead of using the clip-wise video embedding as visual features, we feed embeddings of the whole frames into the model.
The two outputs are combined using a weight $\lambda$.
The parallel cross-attention can be formatted as:

\begin{align}
\begin{split}
\small
Attention(Q, K, V) = softmax&(\frac{QK^{T}_{text}}{\sqrt{d}}) \cdot V_{text} \\
+ \lambda \cdot softmax(\frac{QK^{T}_{frame}}{\sqrt{d}}) &\cdot V_{frame} \text{,}\\
K_{text} = W_{K}^{text} \cdot T_{emb}, V_{text} &= W_{V}^{text} \cdot T_{emb} \text{,}\\
K_{frame} = W_{K}^{frame} \cdot F_{emb}, V_{frame}& = W_{V}^{frame} \cdot F_{emb} \text{,}
\end{split}
\end{align}
where $T_{emb}$ and $F_{emb}$ represent the extracted text embeddings and video frame embeddings, respectively.
During training, only $W_{K}^{frame}$ and $W_{V}^{frame}$ are trainable, enabling a lightweight adaptation to map pre-trained features to the latent space of T2A model inputs.
$W_{K}^{text}$ and $W_{V}^{text}$ are initialized from the pre-trained cross-attention projection layers in Auffusion and kept frozen.
The adapter is trained by the diffusion objective:
\begin{equation}
    \mathcal{L} = \mathbb{E}_{x, \epsilon \sim N(0,1),t,c}  \Vert \epsilon - \epsilon_{\theta}(z_{t}, t, T_{emb}, F_{emb}) \Vert
\end{equation}

\subsection{Temporal Adapter under Label Guidance}
To improve temporal alignment, we incorporate the guidance from textual labels into the temporal condition extraction.
CLIP features of each frame is mapped by a learnable projection layer and cosine similarities are computed between the projected frame embedding and the textual CLIP embedding.
The temporal condition is obtained by binarizing the similarities with a threshold of 0.5.
The temporal conditions enhanced by label guidance enables generation of more temporal-synchronized audio.

The temporal adapter shares the same architecture as the UNet encoder of Auffusion, which follows ControlNet~\cite{zhang2023adding}.
It is trained on AudioSet-strong~\cite{hershey2021benefit}.
During training, the input is the ground truth timestamp condition while the target is the corresponding audio.
Similar to the frame adapter, the temporal adapter is trained by the diffusion loss.
During inference, temporal conditions obtained by CLIP similarities are used, guiding the audio generation.

\subsection{VGGSound-Continuous Filtering}
\begin{figure}
    \centering
    \includegraphics[width=0.95\linewidth]{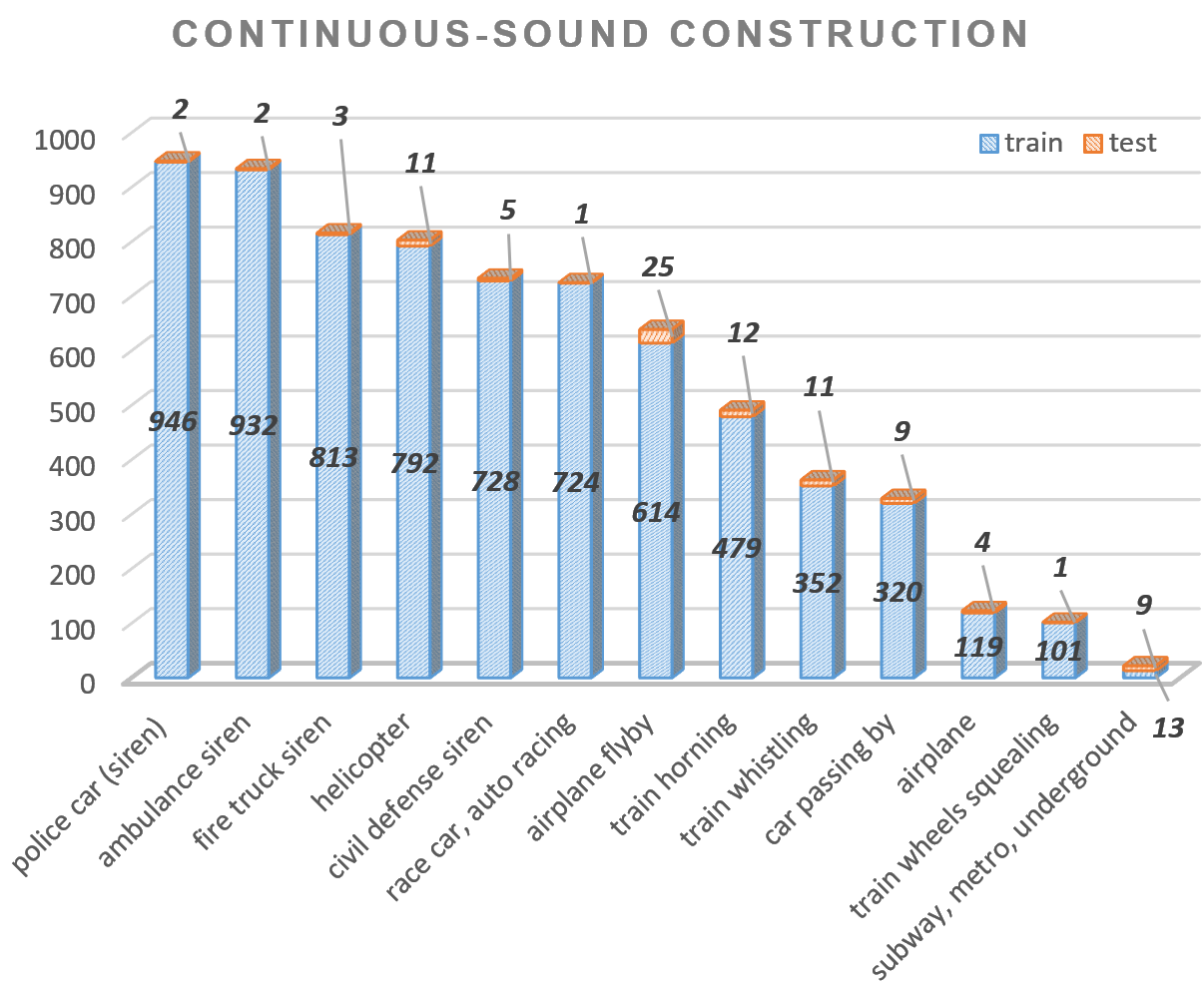}
    \caption{Label distribution of VGGSound-Continuous, most are sounding objects in movement.}
    \label{fig:dataset detail}
\end{figure}

As stated in \Cref{sec:intro}, we find that previous models do not perform well on video with continuous sound.
To improve the generation performance on these video data, we filter the subset with continuous sound, namely \textit{VGGSound-Continuous}, from the commonly-used V2A dataset VGGSound~\cite{chen2020vggsound}.
First, we select out video clips with labels that indicate continuous sound (e.g., siren and airplane sounds).
Then, we use text-to-audio grounding~\cite{xu2024towards} to filter out video clips whose audio do not match their labels.
We manually pick 95 challenging clips as the test split.
The label distribution and statistics are shown in \Cref{fig:dataset detail}.
We initialize Smooth-Foley from FoleyCrafter and fine-tune adapters on VGGSound-Continuous to perform efficient training.

\section{Experiments}
\subsection{Experimental settings}
\subsubsection{Baselines}
We compare Smooth-Foley with two state-of-the-art approaches, Diff-Foley and FoleyCrafter.
Diff-Foley (DF) utilizes contrastive visual-audio pre-trained (CAVP) encoder trained on video-audio pairs to synchronize V2A synthesis.
FoleyCrafter (FC) is the most similar to Smooth-Foley, with the difference that it utilizes clip-wise video embeddings instead of frame-wise ones and it does not incorporate the label for temporal condition extraction.

\subsubsection{Evaluation Metrics}

Following previous works~\cite{luo2024diff,wang2024v2a,zhang2024foleycrafter}, several objective metrics are employed to evaluate the performance of V2A generation, including Frechect Audio Distance~\cite{kilgour19_interspeech} (FAD), Mean KL Divergence~\cite{iashin2021taming} (MKL) and CLIP Score.
FAD models audio embeddings as Gaussian distributions and calculates the distance between generated and ground truth distributions.
FAD based on PANNs~\cite{kong2020panns}, VGGish~\cite{hershey2017cnn} and CLAP~\cite{laionclap2023} are calculated respectively.
MKL measures paired sample-level similarity by calculating the mean KL-divergence across all classes in the test set.
CLIP Score compares the similarity between the input video and the generated audio using Wav2CLIP~\cite{wu2022wav2clip}.
We also perform subjective evaluation, where 10 experienced human evaluators who are familiar with audio generation tasks are invited to rate 10 samples from each model in terms of: 1) semantic alignment; 2) temporal alignment; 3) audio quality. Scores are on a scale of 1 to 10.


\subsubsection{Dataset}
As stated previously, Smooth-Foley is trained on VGGSound-Continuous.
We report evaluation results on the test set of the whole VGGSound and VGGSound-Continuous to compare with previous works.




\begin{table}[h]
\caption{Semantic alignment results on test split of VGGSound-Continuous (VGG-C) and VGGSound (VGG).\\  FC = FoleyCrafter, DF = Diff-Foley.}
\centering
\begin{tabular}{c|cc|cc}
\toprule
\textbf{Method} & \multicolumn{2}{c|}{MKL $\downarrow$}  & \multicolumn{2}{c}{CLIP Score $\uparrow$} \\ 
Dataset & VGG-C  & VGG & VGG-C & VGG \\ 
\hline
FC (w. prom.) & 3.937  & 2.672 & 54.400 & 53.641 \\ 
FC (\textit{w/o} prom.) & 1.855  & 4.032  & 54.400 & 52.811 \\
\hline
DF & 4.844  & 4.770  & 52.867 & 52.779 \\  
\hline
Ours (frame-wise) & \textbf{1.558}  & 2.515  & \textbf{55.124} & \textbf{55.236} \\
Ours (\textit{w/o} frame-wise) & 1.559  & \textbf{2.498}  & 55.076 & 55.233 \\
\bottomrule
\end{tabular}
\label{tab:semantic}
\end{table}
\begin{figure*}[t]
\centering
\includegraphics[width=0.95\textwidth]{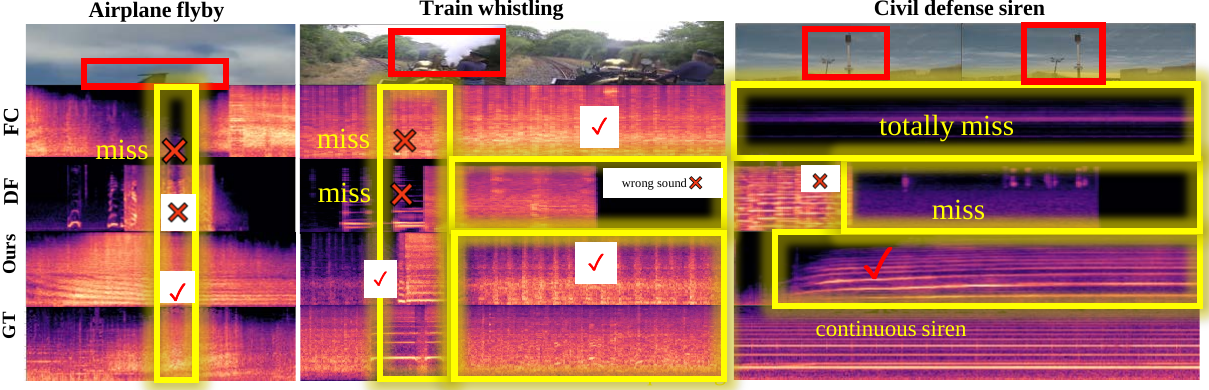}
\caption{Qualitative comparison on temporal alignment with different models, i.e. FoleyCrafter~(FC), Diff-Foley~(DF) and Smooth-Foley (ours).}
\label{fig:qualitative comparison}
\end{figure*}

\begin{table}[h]
\caption{FAD results with different embedding extractor. Numbers on VGG-C / VGG.}
    \centering
\begin{tabular}{c|c c c}
\toprule
\multirow{2}{*}{Model} & \multicolumn{3}{c}{FAD $\downarrow$} \\
   & CLAP & VGGish & PANNs \\ 
\hline
FC (\textit{w.} prompt)& 0.25 / 0.14 & 8.58 / 3.18 & 54.99 / 21.36 \\
FC (\textit{w/o.} prompt) & 0.25 / 0.13 & 8.58 / 5.62 & 54.00 / 29.95 \\
\hline
DF & 0.58 / 0.21 & 28.73 / 6.39 & 106.45 / 34.51 \\
\hline
Ours  & 0.18 / 0.11  & \textbf{5.44} / 2.54 & 33.54 / 18.31 \\
Ours (\textit{w/o.} frame-wise) & 0.18 / \textbf{0.10}  & 5.78 / \textbf{2.26} & \textbf{32.99} / \textbf{13.07} \\
\bottomrule
\end{tabular} 
\label{tab:audio_quality}
\end{table}

    

\begin{table}[h]
\caption{Subjective evaluation results of different models.}
\centering    
\begin{tabular}{c|ccc}
\toprule
Method & Semantic & Temporal & Quality \\
\hline
DF  & 2.44 & 2.58 & 4.27 \\
FC & 5.89 & 5.48 & 5.38 \\
Ours & \textbf{8.42} & \textbf{8.03} & \textbf{6.83} \\
\bottomrule
\end{tabular}
\label{tab:MOS}
\end{table}

\subsection{Comparison with State-of-the-Art}
\subsubsection{Quantitative Comparison}
Quantitative comparison between different models are shown in terms of objective and subjective metrics.
As shown in \Cref{tab:semantic} and \Cref{tab:audio_quality}, on both VGG and VGG-C, Smooth-Foley achieves superior semantic alignment with the video content and provides better audio fidelity.
Under most scenarios, incorporating frame-wise video features achieve better performance than a single clip-wise video feature.
Subjective evaluation results in \Cref{tab:MOS} validates the advantages of Smooth-Foley in semantic alignment and audio quality, while further showing the better temporal alignment of generated audio with the video input.

\subsubsection{Qualitative Comparison}
In \Cref{fig:qualitative comparison}, we list the qualitative comparison results between the ground truth and generation results from different models.
In the first example, when the airplane approaches, FoleyCrafter fails to generate the engine sound.
Diff-Foley produces completely incorrect sounds, but Smooth-Foley successfully generates the corresponding sound effects as the object becomes blurred in the screen.
Additionally, our simulated audios adhere to the Doppler effect principles, demonstrating a rise in frequency as the object moves closer, a peak at the moment of closest encounter, and a subsequent decline.
In contrast, FoleyCrafter exhibits a completely opposite pattern. 
In the second example, multiple events occur.
Train wheels squealing is followed by a steam whistling.
Smooth-Foley generates the steam whistling sound with an accurate onset.
The third example is a nearly static video accompanied by the sound of siren.
FoleyCrafter generates an almost muted audio, completely failing to capture the semantic information from visual frames.
Smooth-Foley successfully produces a high-quality and continuous siren sound.
Diff-Foley fails to correctly generate the sound in all examples.

\subsubsection{Temporal Condition Comparison.}
In \Cref{fig:foleycrafter_deficits}, we pick the top-3 and the bottom-3 estimated probabilities from the time detector of FoleyCrafter and Smooth-Foley.
In continuous video frames, Smooth-Foley constantly follows the visual semantics (in the first case, continuously moving train).
When the main object (in the second case, flying airplane) gradually turns visually ambiguous, the semantic guidance still captures the visual cues and generate correct temporal conditions.

\section{Conclusion}
In this paper, we propose Smooth-Foley to enhance the generation quality of continuous sound using semantic guidance from two aspects.
The first is to replace clip-wise visual embeddings with frame-wise ones, increasing the temporal resolution of visual guidance.
The second is to incorporate the textual label to predict more accurate temporal guidance.
We train a frame adapter and a temporal adapter, which take semantic and temporal conditions respectively, to efficiently adapt a pre-trained T2A model for V2A generation.
We also filter out VGGSound-Continuous, focusing on video with ambiguous sounding object and sound sustainability. 
Based on VGGSound-Continuous, we efficiently enhance pre-trained V2A models for continuous sound generation.
Experiments on VGGSound-Continuous and VGGSound demonstrate that Smooth-Foley achieves superior generation performance against baseline models in terms of audio quality, semantic and temporal alignment.


\newpage

\begin{thebibliography}{10}
\providecommand{\url}[1]{#1}
\csname url@samestyle\endcsname
\providecommand{\newblock}{\relax}
\providecommand{\bibinfo}[2]{#2}
\providecommand{\BIBentrySTDinterwordspacing}{\spaceskip=0pt\relax}
\providecommand{\BIBentryALTinterwordstretchfactor}{4}
\providecommand{\BIBentryALTinterwordspacing}{\spaceskip=\fontdimen2\font plus
\BIBentryALTinterwordstretchfactor\fontdimen3\font minus
  \fontdimen4\font\relax}
\providecommand{\BIBforeignlanguage}[2]{{%
\expandafter\ifx\csname l@#1\endcsname\relax
\typeout{** WARNING: IEEEtran.bst: No hyphenation pattern has been}%
\typeout{** loaded for the language `#1'. Using the pattern for}%
\typeout{** the default language instead.}%
\else
\language=\csname l@#1\endcsname
\fi
#2}}
\providecommand{\BIBdecl}{\relax}
\BIBdecl

\bibitem{chen2020generating}
P.~Chen, Y.~Zhang, M.~Tan, H.~Xiao, D.~Huang, and C.~Gan, ``Generating visually
  aligned sound from videos,'' \emph{IEEE Transactions on Image Processing},
  vol.~29, pp. 8292--8302, 2020.

\bibitem{iashin2021taming}
V.~Iashin and E.~Rahtu, ``Taming visually guided sound generation,'' in
  \emph{British Machine Vision Conference}, 2021.

\bibitem{vaswani2017attention}
A.~Vaswani, ``Attention is all you need,'' \emph{Advances in Neural Information
  Processing Systems}, 2017.

\bibitem{radford2021learning}
A.~Radford, J.~W. Kim, C.~Hallacy, A.~Ramesh, G.~Goh, S.~Agarwal, G.~Sastry,
  A.~Askell, P.~Mishkin, J.~Clark \emph{et~al.}, ``Learning transferable visual
  models from natural language supervision,'' in \emph{International conference
  on machine learning}.\hskip 1em plus 0.5em minus 0.4em\relax PMLR, 2021, pp.
  8748--8763.

\bibitem{girdhar2023imagebind}
R.~Girdhar, A.~El-Nouby, Z.~Liu, M.~Singh, K.~V. Alwala, A.~Joulin, and
  I.~Misra, ``Imagebind: One embedding space to bind them all,'' in
  \emph{Proceedings of the IEEE/CVF Conference on Computer Vision and Pattern
  Recognition}, 2023, pp. 15\,180--15\,190.

\bibitem{ho2020denoising}
J.~Ho, A.~Jain, and P.~Abbeel, ``Denoising diffusion probabilistic models,''
  \emph{Advances in neural information processing systems}, vol.~33, pp.
  6840--6851, 2020.

\bibitem{du2023conditional}
Y.~Du, Z.~Chen, J.~Salamon, B.~Russell, and A.~Owens, ``Conditional generation
  of audio from video via foley analogies,'' in \emph{Proceedings of the
  IEEE/CVF Conference on Computer Vision and Pattern Recognition}, 2023, pp.
  2426--2436.

\bibitem{cui2023varietysound}
C.~Cui, Z.~Zhao, Y.~Ren, J.~Liu, R.~Huang, F.~Chen, Z.~Wang, B.~Huai, and
  F.~Wu, ``Varietysound: Timbre-controllable video to sound generation via
  unsupervised information disentanglement,'' in \emph{ICASSP 2023-2023 IEEE
  International Conference on Acoustics, Speech and Signal Processing
  (ICASSP)}.\hskip 1em plus 0.5em minus 0.4em\relax IEEE, 2023, pp. 1--5.

\bibitem{luo2024diff}
S.~Luo, C.~Yan, C.~Hu, and H.~Zhao, ``Diff-foley: Synchronized video-to-audio
  synthesis with latent diffusion models,'' \emph{Advances in Neural
  Information Processing Systems}, vol.~36, 2024.

\bibitem{sheffer2023hear}
R.~Sheffer and Y.~Adi, ``I hear your true colors: Image guided audio
  generation,'' in \emph{ICASSP 2023-2023 IEEE International Conference on
  Acoustics, Speech and Signal Processing (ICASSP)}.\hskip 1em plus 0.5em minus
  0.4em\relax IEEE, 2023, pp. 1--5.

\bibitem{zhang2024foleycrafter}
Y.~Zhang, Y.~Gu, Y.~Zeng, Z.~Xing, Y.~Wang, Z.~Wu, and K.~Chen, ``Foleycrafter:
  Bring silent videos to life with lifelike and synchronized sounds,''
  \emph{arXiv preprint arXiv:2407.01494}, 2024.

\bibitem{xue2024auffusion}
J.~Xue, Y.~Deng, Y.~Gao, and Y.~Li, ``Auffusion: Leveraging the power of
  diffusion and large language models for text-to-audio generation,''
  \emph{arXiv preprint arXiv:2401.01044}, 2024.

\bibitem{ye2023ip}
H.~Ye, J.~Zhang, S.~Liu, X.~Han, and W.~Yang, ``Ip-adapter: Text compatible
  image prompt adapter for text-to-image diffusion models,'' \emph{arXiv
  preprint arXiv:2308.06721}, 2023.

\bibitem{lee2013pseudo}
D.-H. Lee \emph{et~al.}, ``Pseudo-label: The simple and efficient
  semi-supervised learning method for deep neural networks,'' in \emph{Workshop
  on challenges in representation learning, ICML}, vol.~3, no.~2.\hskip 1em
  plus 0.5em minus 0.4em\relax Atlanta, 2013, p. 896.

\bibitem{lei2015predicting}
J.~Lei~Ba, K.~Swersky, S.~Fidler \emph{et~al.}, ``Predicting deep zero-shot
  convolutional neural networks using textual descriptions,'' in
  \emph{Proceedings of the IEEE international conference on computer vision},
  2015, pp. 4247--4255.

\bibitem{roth2022integrating}
K.~Roth, O.~Vinyals, and Z.~Akata, ``Integrating language guidance into
  vision-based deep metric learning,'' in \emph{Proceedings of the IEEE/CVF
  Conference on Computer Vision and Pattern Recognition}, 2022, pp.
  16\,177--16\,189.

\bibitem{liu2023audioldm}
H.~Liu, Z.~Chen, Y.~Yuan, X.~Mei, X.~Liu, D.~Mandic, W.~Wang, and M.~D.
  Plumbley, ``Audioldm: Text-to-audio generation with latent diffusion
  models,'' \emph{arXiv preprint arXiv:2301.12503}, 2023.

\bibitem{zhang2023adding}
L.~Zhang, A.~Rao, and M.~Agrawala, ``Adding conditional control to
  text-to-image diffusion models,'' in \emph{Proceedings of the IEEE/CVF
  International Conference on Computer Vision}, 2023, pp. 3836--3847.

\bibitem{hershey2021benefit}
S.~Hershey, D.~P. Ellis, E.~Fonseca, A.~Jansen, C.~Liu, R.~C. Moore, and
  M.~Plakal, ``The benefit of temporally-strong labels in audio event
  classification,'' in \emph{ICASSP 2021-2021 IEEE International Conference on
  Acoustics, Speech and Signal Processing (ICASSP)}.\hskip 1em plus 0.5em minus
  0.4em\relax IEEE, 2021, pp. 366--370.

\bibitem{chen2020vggsound}
H.~Chen, W.~Xie, A.~Vedaldi, and A.~Zisserman, ``Vggsound: A large-scale
  audio-visual dataset,'' in \emph{ICASSP 2020-2020 IEEE International
  Conference on Acoustics, Speech and Signal Processing (ICASSP)}.\hskip 1em
  plus 0.5em minus 0.4em\relax IEEE, 2020, pp. 721--725.

\bibitem{xu2021text}
X.~Xu, H.~Dinkel, M.~Wu, and K.~Yu, ``Text-to-audio grounding: Building
  correspondence between captions and sound events,'' in \emph{ICASSP 2021-2021
  IEEE International Conference on Acoustics, Speech and Signal Processing
  (ICASSP)}.\hskip 1em plus 0.5em minus 0.4em\relax IEEE, 2021, pp. 606--610.

\bibitem{xu2024towards}
X.~Xu, Z.~Ma, M.~Wu, and K.~Yu, ``Towards weakly supervised text-to-audio
  grounding,'' \emph{arXiv preprint arXiv:2401.02584}, 2024.

\bibitem{kilgour2018fr}
K.~Kilgour, M.~Zuluaga, D.~Roblek, and M.~Sharifi, ``Fr$\backslash$'echet audio
  distance: A metric for evaluating music enhancement algorithms,'' \emph{arXiv
  preprint arXiv:1812.08466}, 2018.

\bibitem{tailleur2024correlation}
M.~Tailleur, J.~Lee, M.~Lagrange, K.~Choi, L.~M. Heller, K.~Imoto, and
  Y.~Okamoto, ``Correlation of fr$\backslash$'echet audio distance with human
  perception of environmental audio is embedding dependant,'' \emph{arXiv
  preprint arXiv:2403.17508}, 2024.

\bibitem{kong2020panns}
Q.~Kong, Y.~Cao, T.~Iqbal, Y.~Wang, W.~Wang, and M.~D. Plumbley, ``Panns:
  Large-scale pretrained audio neural networks for audio pattern recognition,''
  \emph{IEEE/ACM Transactions on Audio, Speech, and Language Processing},
  vol.~28, pp. 2880--2894, 2020.

\bibitem{laionclap2023}
Y.~Wu*, K.~Chen*, T.~Zhang*, Y.~Hui*, T.~Berg-Kirkpatrick, and S.~Dubnov,
  ``Large-scale contrastive language-audio pretraining with feature fusion and
  keyword-to-caption augmentation,'' in \emph{IEEE International Conference on
  Acoustics, Speech and Signal Processing, ICASSP}, 2023.

\bibitem{hershey2017cnn}
S.~Hershey, S.~Chaudhuri, D.~P. Ellis, J.~F. Gemmeke, A.~Jansen, R.~C. Moore,
  M.~Plakal, D.~Platt, R.~A. Saurous, B.~Seybold \emph{et~al.}, ``Cnn
  architectures for large-scale audio classification,'' in \emph{2017 ieee
  international conference on acoustics, speech and signal processing
  (ICASSP)}.\hskip 1em plus 0.5em minus 0.4em\relax IEEE, 2017, pp. 131--135.

\bibitem{wang2024v2a}
H.~Wang, J.~Ma, S.~Pascual, R.~Cartwright, and W.~Cai, ``V2a-mapper: A
  lightweight solution for vision-to-audio generation by connecting foundation
  models,'' in \emph{Proceedings of the AAAI Conference on Artificial
  Intelligence}, vol.~38, no.~14, 2024, pp. 15\,492--15\,501.

\bibitem{wang2024frieren}
Y.~Wang, W.~Guo, R.~Huang, J.~Huang, Z.~Wang, F.~You, R.~Li, and Z.~Zhao,
  ``Frieren: Efficient video-to-audio generation with rectified flow
  matching,'' \emph{arXiv preprint arXiv:2406.00320}, 2024.

\bibitem{lee2024video}
J.~Lee, J.~Im, D.~Kim, and J.~Nam, ``Video-foley: Two-stage video-to-sound
  generation via temporal event condition for foley sound,'' \emph{arXiv
  preprint arXiv:2408.11915}, 2024.

\bibitem{wu2022wav2clip}
H.-H. Wu, P.~Seetharaman, K.~Kumar, and J.~P. Bello, ``Wav2clip: Learning
  robust audio representations from clip,'' in \emph{IEEE International
  Conference on Acoustics, Speech and Signal Processing (ICASSP)}.\hskip 1em
  plus 0.5em minus 0.4em\relax IEEE, 2022, pp. 4563--4567.

\bibitem{kilgour19_interspeech}
K.~Kilgour, M.~Zuluaga, D.~Roblek, and M.~Sharifi, ``Fréchet audio distance: A
  reference-free metric for evaluating music enhancement algorithms,'' in
  \emph{Interspeech 2019}, pp. 2350--2354.

\end{thebibliography}


\end{document}